\documentclass[prb,aps,twocolumn,draft,tightenlines,%
showpacs,floatfix]{revtex4}
\usepackage{psfig}
\usepackage{amssymb}
\usepackage{amsmath}
\usepackage{colordvi}

\begin{document}

\title{Spin diffusion/transport in $n$-type GaAs quantum  wells}
\author{J. L. Cheng}
\author{M. W. Wu}
\thanks{Author to whom all correspondence should be addressed}
\email{mwwu@ustc.edu.cn.}
\affiliation{Hefei National Laboratory for Physical Sciences at
  Microscale,
  University of Science and Technology of China, Hefei,
  Anhui, 230026, China}
\affiliation{Department of Physics, University of Science and Technology of
  China, Hefei, Anhui, 230026, China}
\altaffiliation{Mailing Address}

\date{\today}

\begin{abstract}
The spin diffusion/transport in $n$-type (001) GaAs quantum well at
high temperatures ($\ge120$\ K) is studied by setting up and
numerically solving the kinetic spin Bloch equations together with
the Poisson equation self-consistently.  All the scattering,
especially the electron-electron Coulomb scattering, is explicitly
included and solved in the theory. This enables us to study the
system far away from the equilibrium, such as the hot-electron
effect induced by the external electric field parallel to the
quantum well. We find that the spin polarization/coherence
oscillates along the transport direction even when there is no
external magnetic field. We show that when the scattering is strong
enough, electron spins with different momentums oscillate in the
same phase which leads to equal transversal spin injection length
and ensemble transversal injection length. It is also shown
that the intrinsic scattering is already strong enough for such a
phenomena. The oscillation period is almost independent on the
external electric field which is in agreement with the latest
experiment in bulk system at very low temperature [Europhys. Lett.
{\bf 75}, 597 (2006)]. The spin relaxation/dephasing along the
diffusion/transport can be well understood by the inhomogeneous
broadening, which is caused by the momentum-dependent diffusion and
the spin-orbit coupling, and the scattering. The scattering,
temperature, quantum well width and external magnetic/electric field
dependence of the spin diffusion is studied in detail.
\end{abstract}

\maketitle

\section{Introduction}
The study of semiconductor spintronics\cite{spintronics}
has caused a lot of attention since long spin lifetime and
large spin transport distance  have been observed in
$n$-type semiconductors,\cite{wolf,kikk1,kikk2,kikk3,hohno,ohno1,kikk4}
due to the great potential of the device applications such as quantum memory
devices, spin valves and spin transistors.
High spin injection efficiency and long spin diffusion/transport distance are
two prerequisites of realizing these devices and have been
widely studied both
experimentally\cite{ohno,waag,awschalom,brand,strand,schmidt,beck,crooker2} and
theoretically.\cite{jonson,taka,oso,flatte,PRB_weng1,jaro,zgyu,martin,JAP_weng,zhang,PRB_weng,JAP_jiang,sakin,osi,PRB_wang,crooker1,sandipan,sham,dan}
Many works are focused on improving the spin injection
efficiency by electrical
 method.\cite{schmidt,strand,rashba1,osi,awschalom,xjiang}
Others are concentrated on the spin diffusion/transport inside the
semiconductors without taking care of the
injection. \cite{PRB_weng1,JAP_weng,zhang,beck,crooker2,JAP_jiang,crooker1,flatte,zgyu,martin,sandipan}
In the latter case, the
spin polarization can be prepared by optical
excitations.\cite{crooker2,beck,meier}
A number of theories have been developed to study the spin transport,
such as the two-component drift-diffusion model,\cite{jaro,zgyu,flatte,martin}
the kinetic spin Bloch equation approach,\cite{PRB_weng1,JAP_weng,JAP_jiang,PRB_weng,crooker1}
 the Monte-Carlo simulation,\cite{sakin,osi,PRB_wang,sandipan,pershin} and the microscopic
semiclassical approach.\cite{saikin,mishchenko,schmeltzer,Bleibaum}
In these theories, Weng and Wu
 showed that the correlations between the spin-up and -down states,
{\em i.e.}, the off-diagonal terms of the density matrix in spin space,
play an essential role in spin diffusion/transport.\cite{PRB_weng1}
By constructing the kinetic spin Bloch equations by means of the non-equilibrium
Green function method and numerically solving these equations with the
scattering explicitly included, they predicted spin oscillations along the
spin diffusion in the absence of the magnetic field\cite{JAP_weng,PRB_weng}
which cannot be obtained from the two-component drift-diffusion model.
These oscillations were later observed in experiments.\cite{beck,crooker2}
Now most of the theoretical works include the off-diagonal
terms.\cite{PRB_weng1,JAP_weng,zhang,PRB_weng,JAP_jiang,sakin,crooker1,sandipan,crooker2,saikin}

Another important consequence of the off-diagonal terms is that
they allow spin to precess along the effective magnetic field
which origins from the spin-orbit coupling (SOC) and
is momentum dependent. The fact that spin with different momentum
precesses with different frequency is referred to
as inhomogeneous broadening.\cite{wu1,Allen}
It was shown by Wu {\em et al.} in the spacial
uniform systems that in the  presence of the
inhomogeneous broadening, any spin conserving scattering, including
the Coulomb scattering, can cause irreversible spin
relaxation/dephasing (R/D).\cite{wu1,wu2,dephasing,hot,zhou,PRB_clv}
Later Weng and Wu extended this concept to the spacial inhomogeneous
systems and showed that the spin with different
momentum precesses with different frequencies during the diffusion
along the spacial gradient, thanks to the off-diagonal term.\cite{PRB_weng1}
This serves as additional inhomogeneous broadening and
can cause additional spin R/D combined with the
spin conserving scattering.\cite{PRB_weng1,JAP_weng,JAP_jiang}

In our previous works, we have studied both the
transient\cite{JAP_weng,JAP_jiang}
and the steady-state\cite{PRB_weng1,JAP_weng,PRB_weng} spin diffusion/transport
in GaAs quantum well (QW) using the kinetic spin Bloch equations,
first without\cite{PRB_weng1,JAP_weng} and
then with\cite{PRB_weng,JAP_jiang} the Coulomb
scattering. The other scattering such as electron-phonon and electron-impurity scattering
is explicitly included. By solving the kinetic spin Bloch equations
combined with the Poisson
equation, one is able to obtain the mobility, diffusion length
and spin injection
length without any fitting parameter. Moreover, this approach is valid not only
for systems near the equilibrium, but also for those
far away from the equilibrium such as systems
under strong external electric field (hot-electron effect)
and/or with large spin polarization. It is also applicable to
systems in both the strong scattering regime and the weak one.\cite{PRB_clv}
Nevertheless, in our previous investigation of the steady-state
spin transport,\cite{PRB_weng1,JAP_weng} the boundary conditions we used
are the single-side ones, {\em i.e.}, $\rho_{k_x\sigma\sigma
^\prime}(x=0)$ are given from the single side of the sample regardless
of the sign of $k_x$, with $\rho_{k_x\sigma\sigma^\prime}(x)$
representing the density matrix elements. This is an approximation as
in principal $\rho_{k_x\sigma\sigma^\prime}(x=0)$ for $k_x<0$
($k_x>0$)  can only be determined from the right
(left) side of $x=0$. Moreover,
by using this boundary conditions, the grid of the space
has to be very small due to the reason specified in the next section.
Therefore  the length of the sample we studied before
is very short, even less than one oscillation period,
as the numerical calculation is too
time-consuming.\cite{PRB_weng1,JAP_weng}  It is therefore also almost
impossible to discuss situations such as spin transport under strong
external electric fields. In this paper, we adopt the
double-side boundary conditions which are  widely used in the
charge transport,\cite{coz1,coz2,shu} {\em i.e.}, using $\rho_{k_x
\sigma\sigma^\prime}(x=0)$ and $\rho_{-k_x\sigma\sigma^\prime}(x=L)$
as boundary conditions with $k_x>0$ and
$L$ standing for the right side of the
sample. We further develop new numerical scheme to solve the kinetic
equations with all the scattering included. This allows us to solve
the spin diffusion/transport with fast speed and high accuracy. We therefore
study the spin diffusion/transport with large sample scale under various
conditions such as temperature, well width, magnetic field and
electric field. The hot-electron effect to the spin transport is
also discussed in detail.

This paper is organized as follows: In Sec.\ II we set up the model
and construct the kinetic spin Bloch equations
with the boundary conditions by using
the Keldysh Green function method.  In Sec.\ III we present our main
results of the spin  diffusion/transport
with different scattering, temperature, quantum
well width, external magnetic field and external electric field.
We conclude in Sec.\ IV. In Appendix\ A we give the
scheme for the numerical calculation.

\section{Model and kinetic spin Bloch equations}
We  begin our study from a two dimensional electron gas (2DEG)
confined by an infinite square well with width $a$
along the (001) direction. The growth
direction is assumed to be along the $z$-axis. A moderate magnetic field
$B$ and an electric field $E$ are applied along the $x$-$y$ plane with the
$x$-axis being the diffusion/transport direction.
The electron state is described
by a subband index $n$ which comes from the
confinement of the QW,  a 2D momentum
 $\mathbf k =(k_x, k_y)$, which represents the momentum
along the $x$-$y$--plane  and the spin index
$\sigma$, which stands for
the spin-up ($\uparrow$) or -down ($\downarrow$) state along
the $z$-axis. In the present paper, the width of the QW is taken to
be so small that only the lowest subband $n=1$ is
occupied and the higher subbands are negligible.

By using the nonequillibrium Green function
method with the gradient expression as well as the generalized
Kadanoff-Baym ansatz,\cite{haug} we construct the kinetic spin Bloch equations
as follows\cite{PRB_weng1,JAP_weng}
\begin{eqnarray}
\frac{\partial}{\partial t} \rho_{\mathbf{k}}(\mathbf{r}, t) &=&
\left.\frac{\partial}{\partial t}\rho_{\mathbf k}(\mathbf{r}, t)
\right|_{\mathtt{dr}}
+ \left.\frac{\partial}{\partial t}\rho_{\mathbf k}(\mathbf{r}, t)
\right|_{\mathtt{dif}}\nonumber\\
&&+ \left.\frac{\partial}{\partial t}\rho_{\mathbf k}(\mathbf{r}, t)
\right|_{\mathtt{coh}}
+ \left.\frac{\partial}{\partial t}\rho_{\mathbf k}(\mathbf{r},t)
\right|_{\mathtt{scat}}\ .
\end{eqnarray}
Here  $\rho_{\mathbf k}(\mathbf
r, t)=\begin{pmatrix}f_{\mathbf k\uparrow} & \rho_{\mathbf
    k\uparrow\downarrow}\\
  \rho_{\mathbf k\downarrow\uparrow} & f_{\mathbf
    k\downarrow}\end{pmatrix}$ are the density matrices of electrons
with momentum ${\bf k}$ at position ${\bf r}=(x,y)$ and time $t$.
The diagonal elements $f_{\mathbf k,\sigma}$ stand for the electron
distribution functions of spin $\sigma$ whereas the
off-diagonal elements $\rho_{\mathbf
  k\uparrow\downarrow}=\rho_{\mathbf k\downarrow\uparrow}^{\ast}
\equiv\rho_{\bf k}$ represent
the correlations between the spin-up and -down states.
\begin{equation}
\label{driving}
\left.\frac{\partial\rho_{\mathbf k}(\mathbf{r}, t)}
{\partial t}\right|_{\mathtt{dr}}=\frac{1}{2}
\{\mathbf{\nabla}_{\mathbf{r}}
\overline{\varepsilon}_{\mathbf{k}}(\mathbf{r}, t),
\mathbf{\nabla}_{\mathbf{k}}\rho_{\mathbf k}(\mathbf{r}, t)\}
\end{equation}
are the driving terms from the external electric field.
Here
\begin{equation}
\overline{\varepsilon}_{\mathbf{k}}(\mathbf{r}, t)=\frac{
{\mathbf  k}^2}{2m^{\ast}}+[g\mu_B{\mathbf B}+{\mathbf h}(\mathbf
k)]\cdot\frac{\mbox{\boldmath$\sigma$\unboldmath}}{2}-e\Psi(\mathbf
r)+{\cal E}_{\mathtt{HF}}({\bf r},t)
\end{equation}
with  $m^{\ast}$ denoting the effective mass.
${\mathbf h}(\mathbf k)$ is the D'yakonov and  Perel' (DP)\cite{dp}
effective magnetic field from the
Dresselhaus\cite{dresselhaus} and the Rashba\cite{rashba} terms.
For GaAs, the Dresselhaus term is the leading term and ${\bf h}({\bf k})$
reads
\begin{equation}
{\bf h}({\bf k})=(\gamma k_x(k_y^2-\langle k_z^2\rangle),
\gamma k_y(\langle k_z^2\rangle-k_x^2), 0)\ .
\label{hk}
\end{equation}
Here the spin splitting parameter $\gamma$ is given by \cite{meier}
\begin{equation}
\gamma=(4/3)(m^{\ast}/m_{cv})(1/\sqrt{2m^{\ast
3}E_g})(\eta/\sqrt{1-\eta/3})\ ,
\end{equation}
in which $\eta=\Delta/(E_g+\Delta)$; $E_g$ denotes the band gap;
$\Delta$ represents the spin-orbit
splitting of the valence band; $m^{\ast}$ standing for the electron mass
in GaAs; and $m_{cv}$ is a constant close in magnitude to the free
electron mass $m_0$.\cite{aronov} For GaAs
$\gamma_0=11.4$\ eV$\cdot$\AA$^3$.  $\langle k_z^2\rangle$
stands for the average of the operator
$-(\partial/\partial z)^2$ over the electronic state of the lowest subband
and being $(\pi/a)^2$ under the infinite-well-depth
assumption. ${\cal E}_{\mathtt{HF}}({\bf r},t)=-\sum_{\mathbf  q}
V_{\mathbf q}\rho_{\mathbf{k-q}}
(\mathbf r,t)$ is the Hartree-Fock term, {\it i.e.} the exchange
term,\cite{haug}  with $V_{\bf q}=\sum_{q_z}\frac{4\pi
e^2}{\kappa_0({\mathbf{q}}^2+q_z^2+\kappa^2)}|I(iq_z)|^2$  the
screened coulomb potential and $\kappa^2=4\pi e^2 n_0/(a k_B T)$
 the Debye-H\"ucke screening constant. Here $\kappa_0$
is the static dielectric constant, $n_0$ represents the
2D electron density and
$|I(iq_z)|^2=\pi^4\sin^2y/[y^2(y^2-\pi^2)^2]$ is the form factor with
$y=q_za/2$. The Hartree-Fock term is important only when there is a large
spin polarization.\cite{dephasing} For the small spin polarization
explored in this paper, it can be ignored. $\Psi(\mathbf r)$ is the electric
potential which  satisfies the Poisson
equation:\cite{possion}
\begin{equation}
{\bf \nabla}_{\mathbf r}^2\Psi(\mathbf r)=e[n(\mathbf r)-N_0(\mathbf
r)]/(a\kappa_0)
\label{eq:possion}
\end{equation}
with $n(\mathbf r)$ standing for the electron density at position $\mathbf r$
and $N_0(\mathbf r)$ representing the background positive charge
density.
The bracket $\{A, B\}=AB+BA$ in Eq.\ (\ref{driving}) is the anti-commutator.
\begin{equation}
\left.\frac{\partial\rho_{\mathbf k}(\mathbf{r}, t)}{\partial
    t}\right|_{\mathtt{dif}}=-{\frac{1}{2}}
\{\mathbf{\nabla}_{\mathbf{k}}\overline
{\varepsilon}_{\mathbf{k}}(\mathbf{r},t),\mathbf{\nabla}_
{\mathbf{r}}\rho_{\mathbf k}(\mathbf{r},t)\}
\end{equation}
are the diffusion terms which also give the additional inhomogeneous broadening
due to the broadening of the spin precession frequencies for different momentum ${\bf k}$
along the spacial gradient.\cite{PRB_weng1}
The coherent terms $\left.\frac{\partial\rho_{\mathbf k}(\mathbf{r}, t)}{\partial
    t}\right|_{\mathtt{coh}}$ and
the scattering terms $ \left.\frac{\partial\rho_{\mathbf k}(\mathbf{r}, t)}{\partial
    t}\right|_{\mathtt{scat}}$ can be found in Refs.\ \onlinecite{hot} and
\onlinecite{helix}.

In the present paper it is assumed that a spin polarization is injected
from the left side of the sample and diffuse/transport along the $x$-axis.
Its spacial distribution along the $y$-axis is uniform.
The kinetic spin Bloch equations are then simplified into
\begin{eqnarray}
\label{eq:BEQ}
&&\hspace{-0.5cm}\frac{\partial \rho_{\mathbf{k}}(x, t)}{\partial t}+
e\frac{\partial \Psi(x,
t)}{\partial x}\frac{\partial \rho_{\mathbf k}(x, t)}{\partial k_x}
 + \frac{k_x}{m^{\ast}}\frac{\partial \rho_{\mathbf k}(x,t)}{\partial x} \nonumber\\
&&\hspace{-0.5cm}+i\left[(g\mu_B\mathbf{B}+\mathbf{h}(\mathbf k))\cdot\frac{\mbox{\boldmath
$\sigma$\unboldmath}}{2}, \rho_{\mathbf k}
(x, t)\right]
= \left.\frac{\partial\rho_{\mathbf k}(x,t)}{\partial
    t}\right|_{\mathtt{scat}}\ .
\end{eqnarray}
Here the spin-orbit coupling $\partial {\bf h}({\bf k})/\partial k_x
\cdot\mbox{\boldmath$\sigma$\unboldmath}/2$
in the diffusion term is very small
in  comparison to $k_x/m^\ast$ and is ignored.
The inhomogeneous broadening can be easily  seen from
these equations. For the spacial uniform case,
$\partial \rho_{\mathbf k}(x, t)/\partial x=0$ and
the electron spin precession is inhomogeneously broadened only
due to the momentum-dependent effective magnetic
field  ${\bf B}_{e}({\bf k})\equiv
{\mathbf h}(\mathbf k)/g\mu_B +
{\bf B}$. With this inhomogeneous broadening, spin
conserving scattering including the Coulomb
 scattering\cite{wu1,PRB_clv,PRB_weng1,ivchenko}
results in an irreversible spin
R/D. Moreover, the scattering can also
give a counter effect to the inhomogeneous broadening.\cite{PRB_clv}
It is noted that unless the $g$-factor is inhomogeneously
broadened,\cite{wu1,ya}  the external magnetic field
itself cannot cause any inhomogeneous broadening. So
there is no spin R/D when ${\bf h}({\bf k})=0$.\cite{wu1}
For the spacial inhomogeneous case, besides the spin precession
around the effective magnetic field, its diffusion into the semiconductor
also accompanies a spin precession with the precession frequency
being $\mathbf k$-dependent, {\em i.e.}, the
third term in Eq.\ (\ref{eq:BEQ}). So this causes additional
inhomogeneous broadening. Moreover, in the steady-state spin
diffusion/transport, the spin precession rate along the diffusion
direction is determined by $|g\mu_B{\bf B}+{\bf h}({\bf k})|/k_x$
according to Eq.\ (\ref{eq:BEQ}). Therefore,
differing from the spacial uniform case,
the external magnetic field itself {\em alone} can
cause spin R/D along the diffusion direction due to the
inhomogeneous broadening of the diffusion
velocity $k_x/m^\ast$.\cite{PRB_weng1}

The kinetic spin Bloch equations\ (\ref{eq:BEQ}) together with the Poisson
equation\ (\ref{eq:possion}) are highly nonlinear
and have to be solved numerically. In our previous
paper,\cite{JAP_weng} this set of equations are solved with
single-side boundary conditions.
In such a method, an extra parameter, so called the
coefficient of viscosity,\cite{coz1}
 is introduced  to keep the smoothness of the
solution and the Poisson equation is inevitable to keep the system
from divergence even though there are no charge imbalance and
external electric field. This can be seen by retaining the third terms
in Eq.\ (\ref{eq:BEQ}) and replacing the scattering terms by the
relaxation time approximation:
\begin{equation}
\frac{k_x}{m^{\ast}}\frac{\partial \rho_{\mathbf k}(x)}{\partial x}
=-\frac{\rho_{\bf k}(x)-\rho_{\bf k}^0}{\tau_p}\ ,
\label{div}
\end{equation}
with $\rho_{\bf k}^0$ representing the density matrices in the equilibrium.
It is clearly seen from this equation that $\rho_{\bf k}(x)$ is divergent
when $x\rightarrow \infty$  for $k_x<0$. In order to circumvent this
divergence, the Poisson equation has to be applied and the grid for the
space has to be very small.
In the present investigation we extend the conventional numerical scheme of
solving the Boltzmann equation for charge transport\cite{coz1,coz2} to
solve the kinetic spin Bloch equations by using the
double-side boundary conditions. This conditions assume a finite
sample length $L$ and are given by
\begin{eqnarray}
\begin{cases}
  \rho_{\mathbf k}(x=0, t) = F(\varepsilon_{\mathbf k,\mathtt{E}}-\mu(0))\ ,
 & \mbox{for}\ k_x>0\ , \\
  \rho_{\mathbf k}(x=L, t) = F(\varepsilon_{\mathbf k,\mathtt{E}}-\mu(L))\ ,
 & \mbox{for}\ k_x<0\ .
\end{cases}
\label{eq:boundary}
\end{eqnarray}
Here $F(\varepsilon_{\mathbf k,\mathtt{E}}-\mu(x))\equiv
[\exp(\varepsilon_{\mathbf k,\mathtt{E}}-\mu(x))/(k_BT_e)+1]^{-1}$
represent the
drifted Fermi distributions in the spin space,
with the chemical potential matrices
$\mu(0)=\begin{pmatrix}\mu_{\uparrow}&0 \\
  0&\mu_{\downarrow}\end{pmatrix}$
and $\mu(L)=\begin{pmatrix}\mu_0 & 0 \\
  0&\mu_0\end{pmatrix}$ if the spin injection is from the left side and
$L$ is assumed to be large enough so that the spin polarization vanishes
before it reaches the right boundary.
$\varepsilon_{\mathbf k,\mathtt{E}}=(\mathbf
k-m^{\ast}\mu_{\mathtt{E}}\mathbf E)^2/2m^{\ast}$ is energy spectrum
drifted by the electric field.  $T_e$ and $\mu_{\mathtt{E}}$ are
the hot electron temperature and mobility respectively
which are calculated at the spacial uniform case following the
method addressed in our previous paper.\cite{hot}
The initial electron distributions are given by
\begin{eqnarray}
  \rho_{\mathbf k}(x, t=0) = F(\varepsilon_{\mathbf k,\mathtt{E}}-\mu(x))\ ,
\end{eqnarray}
with $\mu(0)=\begin{pmatrix}\mu_{\uparrow}&0 \\
  0&\mu_{\downarrow}\end{pmatrix}$
and $\mu(x\neq0)=\begin{pmatrix}\mu_0 & 0 \\
  0&\mu_0\end{pmatrix}$.
The chemical potentials
 $\mu_{\uparrow(\downarrow)}$ and $\mu_0$
are determined by the equations
\begin{eqnarray}
&&\sum_{\mathbf k}\mbox{Tr}[\rho_{\mathbf k}(x, 0)] = n(x)\ ,\\
&&\sum_{\mathbf k}\mbox{Tr}[\rho_{\mathbf k}(0, 0)\sigma_z]=Pn(0)\ ,
\label{P}
\end{eqnarray}
with $P$ denoting the spin polarization at the boundary $x=0$. The
double-side
boundary conditions are in good consistence with the
physical consideration that the electron must move along its
momentum direction.
For the Poisson equation, its boundary conditions are set to be $\Psi(0)=0$
and $\Psi(L)=EL$ with $E$ representing the strength of the external
electric field.
The numerical scheme for the scattering terms
has been laid out in detail in our previous paper\cite{hot} and
the new numerical scheme for solving the kinetic spin
Bloch equations is given in Appendix\ \ref{method}.
By using this new scheme, we further remove the extra
parameter---the  coefficient of viscosity.

\section{Numerical Results}
We numerically solve the kinetic spin Bloch equations\ (\ref{eq:BEQ})
together with the Poisson equation\ (\ref{eq:possion}) iteratively
to achieve the self-consistent solution.  We include all the
scattering, {\em i.e.}, the electron-electron Coulomb, the
electron-phonon  and the electron-impurity scattering. As we are
interested in the diffusion/transport properties at high
temperatures ($T\ge120$\ K), for the electron-phonon
scattering we only need to consider the
electron--longitudinal-optical (LO) phonon
scattering.\cite{Yu,conwell,manhan} All matrix
elements of the interactions and the expressions of the scattering
terms can be found in Ref.\ \onlinecite{hot}. The width of the QW
and the electron density are taken to be $a=7.5$\ nm and
$N_e=4\times10^{11}$\ cm$^{-2}$ separately unless otherwise
specified. The initial spin polarization of the left boundary is
$P=5\%$. In the numerical calculation, we divide the momentum ${\bf
k}$-space into $N\times M$ control regions as plotted in Fig.\ 8 in
Ref.\ \onlinecite{hot} and the corresponding center points are
\begin{equation}
\mathbf k_{n,m}=\sqrt{(n+\frac{1}{2})\Delta E}
\left(\sin\frac{2m\pi}{M},
  \cos\frac{2m\pi}{M}\right)
\label{kmn}
\end{equation}
 with $n=0,\cdots ,N-1$ and
$m=0,\cdots ,M-1$. In the present work, we take
$(N, M)=(16,18)$.\cite{converg}
The energy cut is determined by the ratio between the distribution at the
energy cut and the distribution at the zero energy is less than 0.1\ \%,
 which is around $10E_f$ at $T=200$\ K with
$E_f=14.3$\ meV being the Fermi energy.
The sample length is usually taken to be $L=10$\
 $\mu$m, unless otherwise specified.
This length has to be taken long enough so that the spin polarization
effectively  vanishes at the right boundary. The other parameters
are listed in Table\ I.\cite{para}
\begin{table}[h]
\begin{tabular}{lllll}
\hline\hline
    $\kappa_0$ & 10.8 &\hspace{2pt}& $\kappa_{\infty}$ & 12.9\\
    $\Omega_{\mathtt{LO}}$ & 35.4 meV && $m^{\ast}$ & 0.067$m_0$\\
    $g$ & 0.44 && $E_g$ & 1.55 eV \\
    $\alpha_0$& 5.33${\AA}$ && $\Delta$ & 0.341 eV \\
\hline\hline
  \end{tabular}
\caption{Parameters used in the numerical calculations,
$\Omega_{\mathtt{LO}}$ is the LO phonon energy.}
\end{table}

Three quantities are used to describe the spin diffusion/transport lengths in the
steady state:
spin-dependent electron density
\begin{equation}
N_{\sigma}(x)=\sum_{\mathbf{k}}f_{\mathbf k,\sigma}(x)\ ,
\label{eq:defN}
\end{equation}
the incoherently summed spin coherence
\begin{equation}
\rho(x)=\sum_{\mathbf{k}}|\rho_{{\bf
        k}\uparrow\downarrow}(x)|\ ,
\label{eq:defrho}
\end{equation}
and the coherently summed spin coherence
  \begin{equation}
\rho^{\prime}(x)=|\sum_{\mathbf{k}}\rho_{{\bf
        k}\uparrow\downarrow}(x)|\ .
    \label{eq:defrho}
\end{equation}
Unlike the corresponding quantities defined in the time domain,\cite{clvwu}
  these three quantities are defined in the spacial domain.
The irreversible spin dephasing length $L_p$
can be deduced from the decay of the envelope of $\rho$;\cite{wu,kuhn,PRB_weng1}
the ensemble spin diffusion length $L_p^{\ast}$
  is given by the decay of the envelope of $\rho^{\prime}$;
whereas the spin diffusion length $L_d$ is
determined from the decay of the envelope of $\Delta
N=N_\uparrow-N_\downarrow$.

\subsection{Transient momentum-resolved spacial evolution of spin polarization}

\begin{figure}[htp]
\centerline{\psfig{file=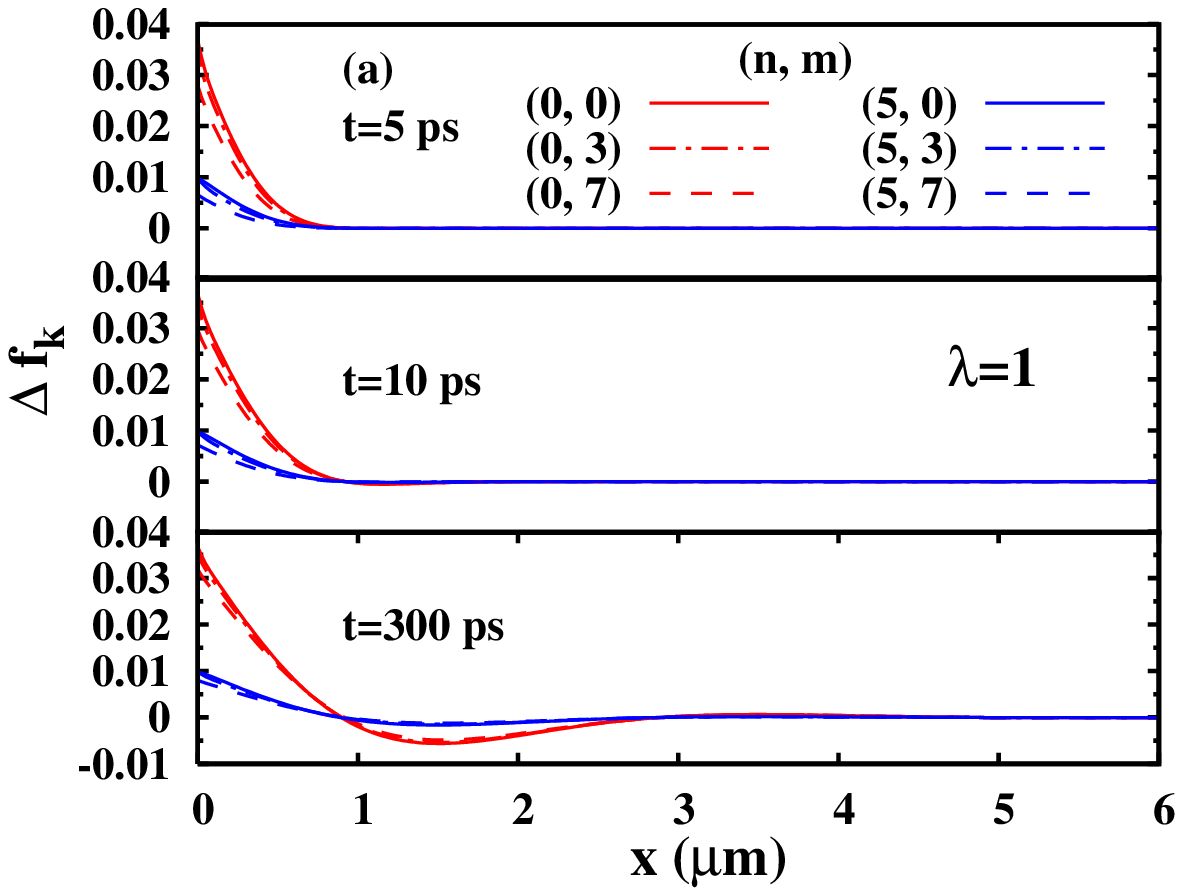,width=7.cm}}
\centerline{\psfig{file=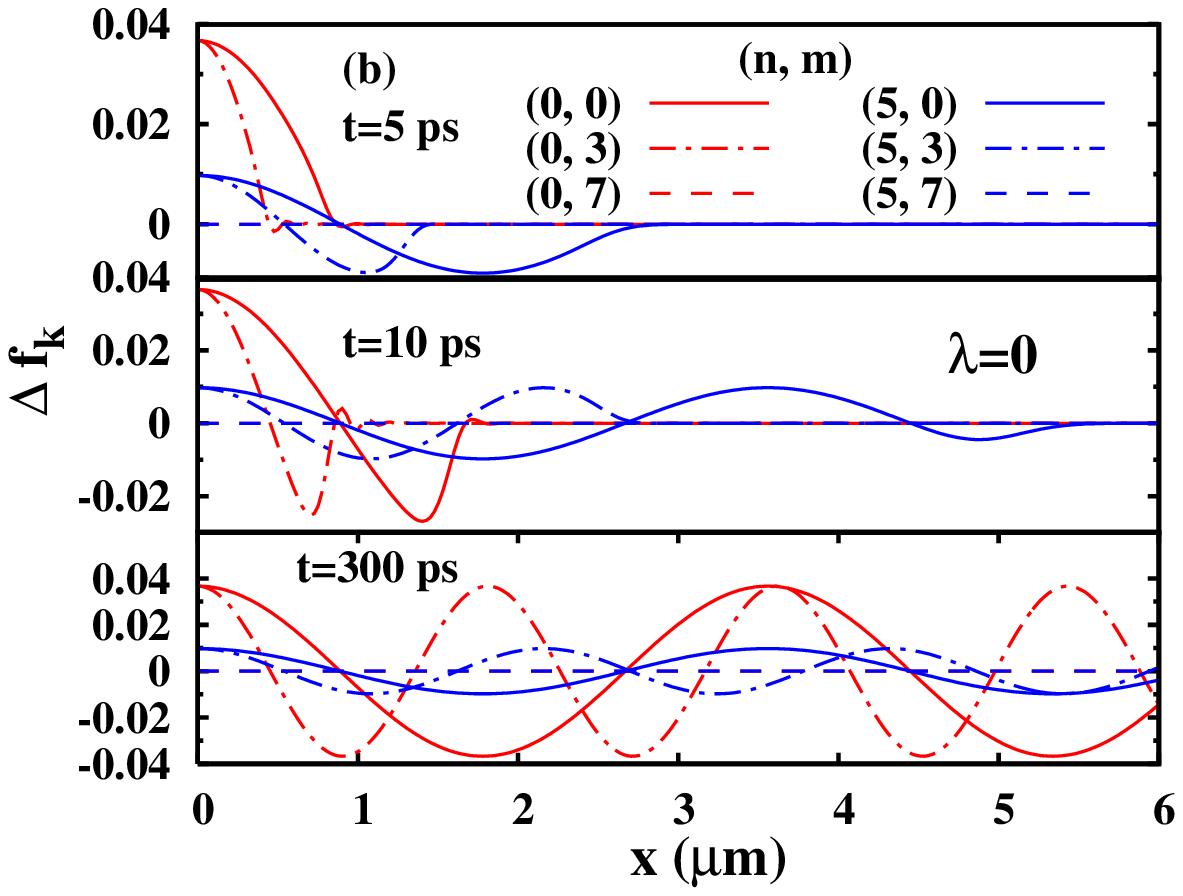,width=7.cm}}
\centerline{\psfig{file=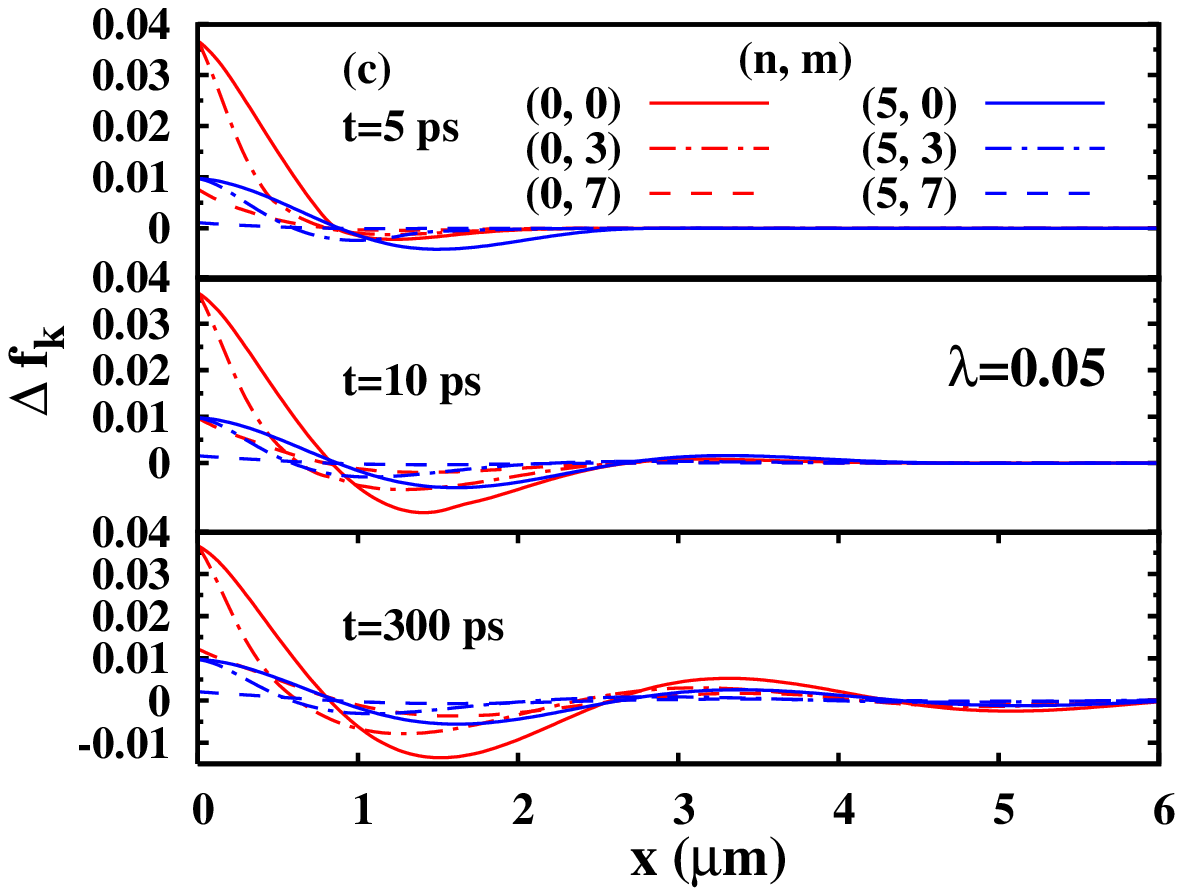,width=7.cm}}
\centerline{\psfig{file=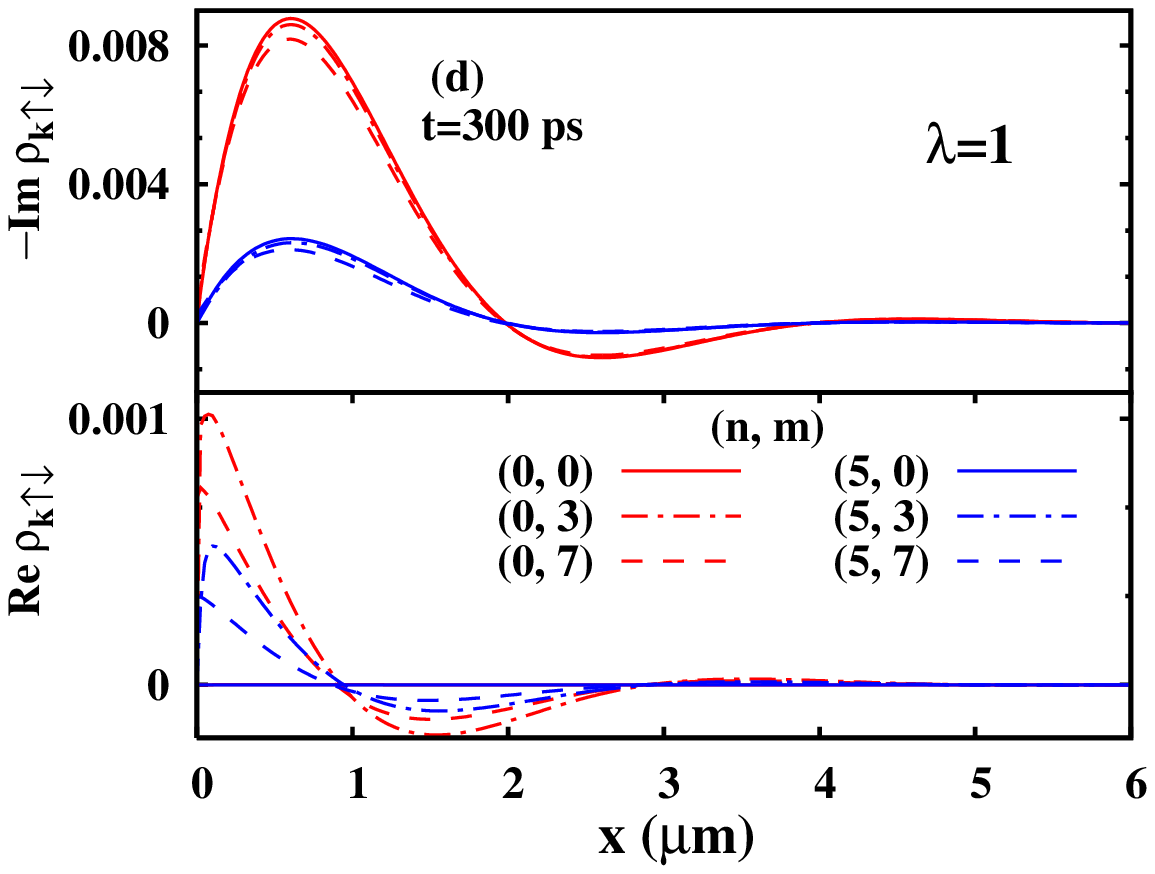,width=7.cm}}
  \caption{(Color online)  $\Delta f_{\mathbf k}$ [(a)-(c)]
and $\rho_{{\bf k}\uparrow\downarrow}$ [(d)] {\em vs.} $x$
for different ${\bf k}_{n,m}$ at  $t=5$, 10  and 300\ ps
(with the system being in the  steady state at 300\ ps).
$T=200$\ K and $B=0$.
(b) no scattering: $\lambda=0$; (c) small scattering:
$\lambda=0.05$; (a) and (d) normal scattering $\lambda=1$.
Only the electron-electron and the electron-LO phonon scattering
is included in the calculation.
It is noted that although $x$ is plotted up to 6\ $\mu$m,
the sample  length $L$ used in the calculation is 10\ $\mu$m.
}
  \label{fig:distribution}
\end{figure}

We first show the spacial dependence of $f_{\mathbf k,\sigma}$ for
different momentums $\mathbf k$ at $t=5$, 10 and 300\ ps in
Fig.\ \ref{fig:distribution}(a) with the electron-electron and
the electron--LO-phonon scattering included.  $T=200$\ K and $B=0$.
Differing from the intuition that the spin precession should
be controlled by the effective magnetic field $\mathbf h(\mathbf k)$,
it is seen from the figure that the electron spins
with different momentums present almost the identical precession
period. This is understood due to the strong scattering which enforces
different spins to precess with the same frequency.
In order to reveal this effect, we scale all the scattering terms
with a scaling factor $\lambda$, {\em i.e.}, replacing the scattering
terms  in Eq.\ (\ref{eq:BEQ}) by
$\lambda\frac{\partial \rho_{\bf k}(x,t)}{\partial t}|_{\mbox{scat}}$.
First we consider the case without any scattering and
electric field, {\em i.e.},
$\lambda=0$ and ${\bf E}=\nabla_{\bf r}\Psi({\bf r})=0$.
It is then  easy to obtain the analytical solution of
Eq.\ (\ref{eq:BEQ}) in the
 steady state with the boundary conditions (\ref{eq:boundary}):
\begin{widetext}
\begin{equation}
  \label{eq:noscat}
\rho_{\mathbf k}(x)
=\begin{cases}e^{-\frac{im^{\ast}}{2k_x}g\mu_B{\bf B}_{e}({\bf k})
  \cdot\mbox{\boldmath$\sigma$\unboldmath}x}
  \rho_{\mathbf k}(x=0)e^{\frac{im^{\ast}}{2k_x}g\mu_B{\bf B}_{e}({\bf k})\cdot
  \mbox{\boldmath$\sigma$\unboldmath}x},\hspace{1cm} k_x>0
  \\ \rho_{\mathbf k}(x=L),\hspace{6.1cm} k_x<0\end{cases}
\end{equation}
\end{widetext}
by further neglecting the HF terms.
This solution  is somehow different from our previous analytical
result\cite{JAP_weng} due to the different boundary conditions.
It describes the injection of the electron spin  with
momentum $\mathbf k$ from both boundaries into the 2DEG.
For each electron with $k_x<0$, it comes from the right boundary ($x=L$)
where there is no spin polarization. So there is no spin precession in the
2DEG. However, for electron with $k_x>0$, it comes from the left boundary
where the electron spin is polarized. The spin polarization precesses
in the 2DEG due to the presence of the effective magnetic
field ${\bf B}_e(\mathbf k)$. Moreover it does not decay in the
absence of any scattering. It is noted from Eq.\ (\ref{eq:noscat})
that unlike the temporal spin precession where the precession
rate is determined by $g\mu_B|{\bf B}_e({\bf k})|$, the spacial spin
precession rate is determined by
\begin{equation}
\label{omega}
\mathbf \Omega_{\bf k}=
m^{\ast}g\mu_B\mathbf B_e(\mathbf k)/k_x
\end{equation}
which strongly depends on the velocity $k_x/m^\ast$. This is
 because in Eq.\ (\ref{eq:BEQ})
the diffusion term is proportional to the velocity.
This gives strong inhomogeneous broadening along the spin
diffusion.\cite{PRB_weng1}

Then we introduce the scattering into the system, first a small
one with $\lambda=0.05$ [Fig.\ \ref{fig:distribution}(c)]
and then  the normal one with
$\lambda=1$ [Fig.\ \ref{fig:distribution}(a)], which are compared with the
case with $\lambda=0$ [Fig.\ \ref{fig:distribution}(b)].
Once the scattering term is turned on, the
electron spin presents a very different evolution behavior from
Eq.\ (\ref{eq:noscat}).
First, electrons with $k_x<0$ [curves for different wavevector
${\mathbf k}_{n,m}$ (Eq.\ (\ref{kmn})) with $(n,m)=(0,7)$ and
$(5,7)$] obtain a spin polarization due to the scattering
which scatters electrons with $k_x>0$ to those with $k_x<0$ and
hence the spin polarization is transferred from electrons with $k_x>0$.
Consequently  this kind of spin polarization is also
built up from the left edge to the
right one even though $k_x<0$.
Second, the scattering tends to drive the spin precession and spin
diffusion to reach the same precession frequency and diffusion velocity
respectively for different ${\bf k}$. This can be seen from
Fig.\ \ref{fig:distribution} where $\Delta f_{\bf k}=f_{{\bf k}\uparrow}-
f_{{\bf k}\downarrow}$ is plotted against the position $x$ for different
momentum ${\bf k}$ at $t=5$, 10, and 300\ ps (already in the steady state)
by comparing the results with $\lambda=0$ (no scattering),
$\lambda=0.05$ (weak scattering) and $\lambda=1$
(normal scattering). In the computation we only include the
electron-electron and electron--LO-phonon scattering. It is easy to find that
when there is no/weak scattering, spin precession frequencies and
spin diffusion velocities with different momentum
show different values. However,
when the scattering becomes stronger,  the
spacial  spin precession frequencies and the spin diffusion velocities
for different momentum ${\bf k}$ tend to a single frequency and velocity as
shown in Fig.\ \ref{fig:distribution}(a). The same behavior is also observed
for the spin coherence $\rho_{{\bf k}\uparrow\downarrow}$ as shown in
Fig.\ \ref{fig:distribution}(d) in the steady state ($t=300$\ ps).
This is also the reason that the numerical results are highly
accurate even for small grid numbers in the $\mathbf k$-space.
Moreover, it shows that
when the scattering is strong enough, there is no interference-induced
decay and consequently $L_p=L_p^{\ast}$. The similar results
have been obtained in the time domain.\cite{clvwu} In fact, even for the case
$N_e=4\times 10^{10}$\ cm$^{-2}$ and $T=120$\ K,
the intrinsic scattering is already strong
enough to ensure this behavior. Spin echo experiments are needed to verify our
findings here.\cite{echo}

This behavior can be understood as following:
Without the scattering, each spin precesses with its own
frequency ${\bf\Omega}_{\bf k}$ independently with the ${\bf k}$-dependent
precession frequency referred to as the inhomogeneous broadening.
Scattering tends to suppress this inhomogeneous broadening and to make it a
homogeneous one. Our result indicates that the intrinsic scattering (the
electron-electron and electron-phonon scattering) is already
strong enough to totally suppress the inhomogeneous
broadening. The spin polarizations for different ${\bf k}$ all precess
with single frequency $\langle{\bf \Omega}_{\bf k}\rangle={\bf
\Omega}_{0}=m^{\ast}\gamma(\langle
k_y^2\rangle -\langle k_z^2\rangle, 0, 0)$ when $B=0$.
In the case we calculate here,
$\Omega_{0}=-1.61\mbox{/}\mu\mbox{m}$ ($\gamma=11.4$\ eV$\cdot$\AA$^3$,
$\langle k_y^2\rangle\approx\frac{1}{2}\frac{2m^{\ast}}{\hbar^2}k_BT$ for the
non-degenerate case) which corresponds to the spacial period $3.88$\ $\mu$m.

Moreover, it is seen from the figure that besides the counter effect of the
scattering to the inhomogeneous broadening, it also causes irreversible
spin R/D.\cite{wu1} In the steady state, there is no spin relaxation for each
${\bf k}$, as shown in Fig.\ \ref{fig:distribution}(b)
in the absence of any scattering.
Whereas strong spin relaxation is observed in Fig.\ \ref{fig:distribution}(a)
in the presence of unscaled  intrinsic scattering.

Finally the scattering also
affects the total spin polarization at the left boundary in the steady state.
It is noted that the boundary conditions Eq.\ (\ref{eq:boundary})
at $x=0$ are only for $k_x>0$. The total spin polarization
Eq.\ (\ref{P}) is determined by the contributions from both $k_x>0$
and $k_x<0$. Without scattering, the spin polarization
for $k_x<0$ is always zero [Eq.\ (\ref{eq:noscat}) and
Fig.\ \ref{fig:distribution}(b)]. Therefore, the total spin polarization
is only one half of 5\ \% at the left boundary. The scattering transfers
spin polarization from states of  $k_x>0$. Therefore, the larger the
scattering, the closer the total spin polarization approaches the boundary
condition: $P=3.0$\ \% for $\lambda=0.05$
and $4.5$\ \% for $\lambda=1.0$.
We believe that the spin polarization at the left boundary can
be very close to $5$\ \% for sufficiently strong scattering. In such a
case, the spin injection with double-side boundary conditions can be
approximated by the single-side boundary conditions used in our previous
papers.\cite{PRB_weng1,JAP_weng} However,
for small spin polarizations, the spin injection properties
such as the spin diffusion length and the spin precession
period are not determined by the initial spin polarization at the boundary.

\subsection{Effect of scattering on spin diffusion}
\begin{figure}[htp]
\centerline{\psfig{file=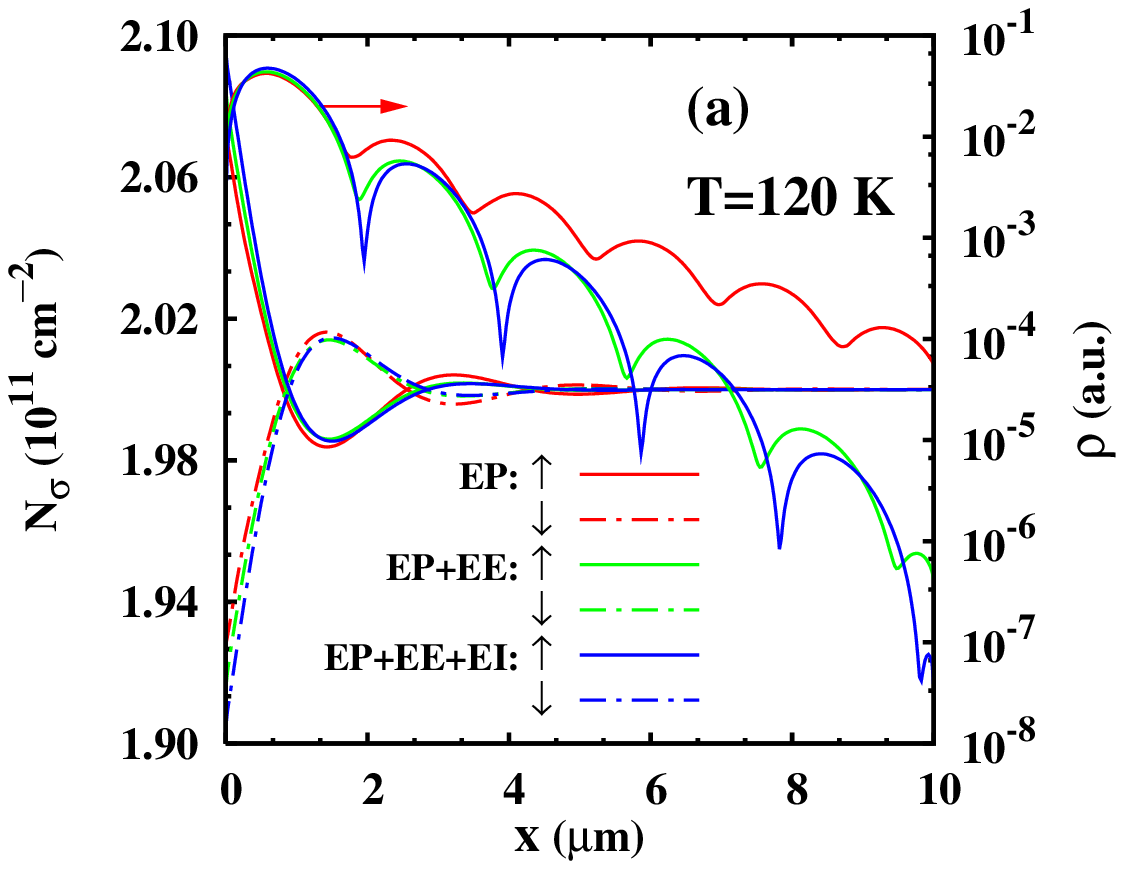,width=7.cm}}
\centerline{\psfig{file=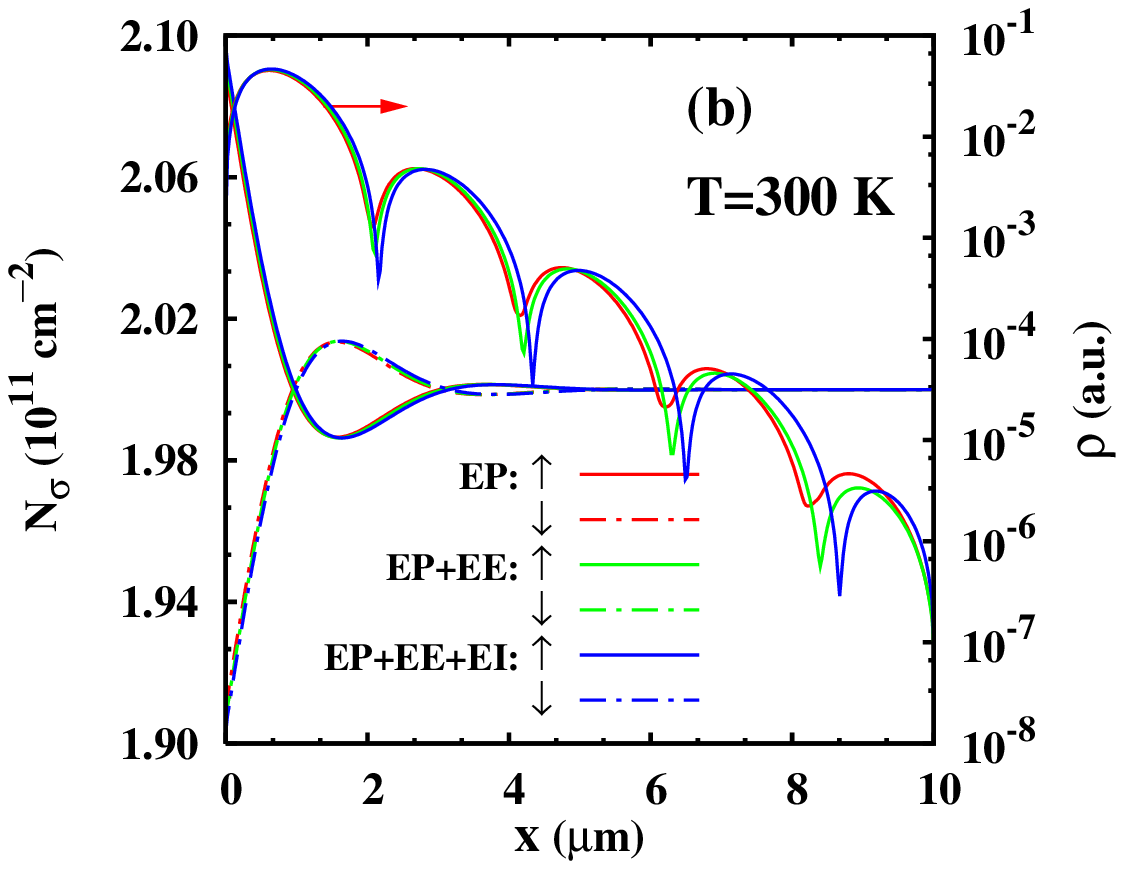,width=7.cm}}
\caption{(Color online) Effect of the scattering on spin diffusion
in the steady state at (a) $T=120$\ K and (b) $T=300$\ K.
Red curves: with only the electron--LO-phonon (EP) scattering;
Green curves: with both the
  electron-electron (EE) and EP scattering; Blue curves: with
  all the scattering, {\em i.e.}, the EE, EP and
  electron-impurity (EI) scattering. The impurity density $N_i=N_e$. Note the scale of the
incoherently summed spin coherence is on the right hand side of the figure.}
\label{fig:scat}
\end{figure}

As shown in the previous subsection that the scattering plays a
crucial role when the electron diffuse/transport into the 2DEG,
we now study the effect of the scattering on the total spin signal.
In Fig.\ \ref{fig:scat} the spin-resolved electron density $N_\sigma$
[Eq.\ ($\ref{eq:defN}$)] and the incoherently summed spin
coherence $\rho$ [Eq.\ ($\ref{eq:defrho}$)]
in the steady state  are plotted  against the position $x$
by first including  only the electron--LO-phonon scattering (red curves),
 then adding the electron-electron scattering (green curves) and
finally adding the electron-impurity scattering (blue curves) with
$N_i=N_e$ at $T=120$\ K (a) and 300\ K (b). $B=0$ in the calculation.

It is seen from the figure that
similar to the transient spin diffusion in our previous
investigation,\cite{JAP_weng,PRB_weng,JAP_jiang}  both $N_{\sigma}$ and $\rho$ oscillate
with the position even when there is {\em no} external magnetic field here.
Besides the oscillation, strong
spin relaxation and dephasing are observed.
Nevertheless, they present a more
complicated behavior: When $T=120$\ K, the spin relaxation and
dephasing always decrease when a new
scattering is added. However, when $T=300$\ K,  they become
faster when the electron-electron scattering is added to the case
with only the electron--LO-phonon scattering but become slower
when electron-impurity scattering is further added.
Finally, the spin polarizations at
the left boundary are  4.75\ \%, 4.20\ \% and
3.60\ \% at 120\ K and
4.8\ \%, 4.7\ \% and 4.6\ \% at 300\ K  for the cases with
all the scattering, with both the electron-electron and the electron--LO-phonon scattering
and with only the electron--LO-phonon scattering, respectively.
{\em i.e.}, the spin polarization at
the left boundary always increases with the strengthening of the scattering,
either the strengthening is due to the addition of more
 scattering or due to the
increase of temperature. This is consistent with the result in
the previous subsection.

\begin{figure}
\centerline{\psfig{file=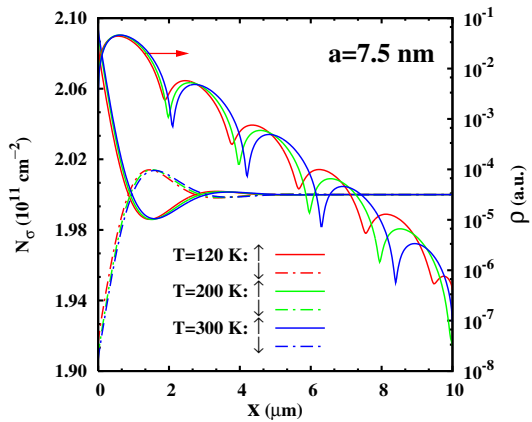,width=7.cm}}
\caption{(Color Online)  $N_{\sigma}$ and  $\rho$
  {\it vs.} the position $x$ at
  $T=120$, 200 and 300\ K with width $a=7.5$\ nm and $L=10$\ $\mu$m.
$N_i=0$.
 Note the scale of the
incoherently summed spin coherence is on the right hand side of the figure.}
\label{fig:tempt}
\end{figure}

The spin precession in the steady-state spin injection in the absence of
the magnetic field is in consistence with the latest experimental observation
in bulk system.\cite{beck}
This kind of spacial precession is different from the spin precession in the
time domain,\cite{brand} where the spin precession can also be observed
in the absence of any magnetic field in 2DEG when
$T<2$\ K. This is because the inhomogeneous broadening in the time domain is
${\bf h}({\bf k})$ with its average over a homogenous electron
distribution being $\langle{\bf h}({\bf k})\rangle=0$,
 which is too weak to sustain the scattering when the temperature
is higher than 2\ K. However, the spin precession in the real space during the diffusion
is quite different. Here the inhomogeneous broadening is caused
by ${\bf \Omega}_{\bf k}=m^\ast{\bf h}({\bf k})/k_x$
with $\langle{\bf \Omega}_{\bf k}
\rangle={\bf \Omega}_0\not=0$.  From previous subsection, we know
the precessions of different wave vector $\mathbf k$ share the same
frequency $\Omega_0$ even with the normal intrinsic scattering.
So the total spin polarization oscillates in the
space and this kind of oscillations exist even in the room temperature.
However, in the experiments by Beck {\em et al.},\cite{beck}
as $\langle k_y^2\rangle=
\langle k_z^2\rangle$ in the bulk system, $\Omega_0=0$ and the effective
magnetic field is provided by strain which is very small. This is the
reason in bulk one has to
observe the spin oscillations at very low temperature.

The complicated scattering dependence of the spin R/D
in the real space is similar to that of the spin R/D in
the time domain,\cite{zhou,PRB_clv} where the scattering dependence of the
spin R/D time is different in the strong and weak scattering
regimes.\cite{PRB_clv} In the weak scattering limit, adding a
new scattering increases the spin R/D as it provides more spin
R/D channel in the presence of the inhomogeneous broadening.
Nevertheless, in the strong scattering
limit, the counter effect of the scattering to the inhomogeneous broadening
is important and adding a new scattering leads to a weaker spin
R/D.\cite{PRB_clv} Similarly,
the spin diffusion length, which is used to represent the
spin relaxation in the real space, is also controlled
by the inhomogeneous broadening and the scattering, with the relative
strength of them dividing the system into the weak/strong scattering
regime. At $120$\ K, the electron--LO-phonon
scattering is very weak and falls into the weak scattering
regime. Therefore the diffusion length decreases with the scattering.
But at  $300$\ K, the scattering becomes stronger. Our results indicate
the system is near the transition regime between the strong and weak
scattering and therefore the changes are marginal by
adding new scattering.

\subsection{Temperature and well width dependence of spin diffusion}

In Fig.\ \ref{fig:tempt} the steady-state spin-resolved electron density
$N_{\sigma}$ and the incoherently summed spin coherence $\rho$ are
plotted as functions of position $x$ at different temperatures $T=120$,
200 and 300\ K for QW with $a=7.5$\ nm.
In the calculation only the intrinsic scattering is included.
One finds from the figure that the spin precession period
increases with the temperature. This is because the period is around
$|\Omega_0^{-1}|=[m^{\ast}\gamma|\langle
k_y^2\rangle -(\pi/a)^2|]^{-1}$. With the increase of temperature,
$\langle k_y^2\rangle$ increases. Consequently the period
of the spacial spin oscillation becomes longer.

The spin diffusion length decreases with the temperature.
This is understood that with the increase of temperature, both
the inhomogeneous broadening and the scattering increase. In the
weak scattering regime, this increase leads to a faster
spin R/D. Nevertheless, the temperature dependence of the spin diffusion
length is very mild compared to the temperature dependence of
the spin precession period.

\begin{figure}[htp]
\centerline{\psfig{file=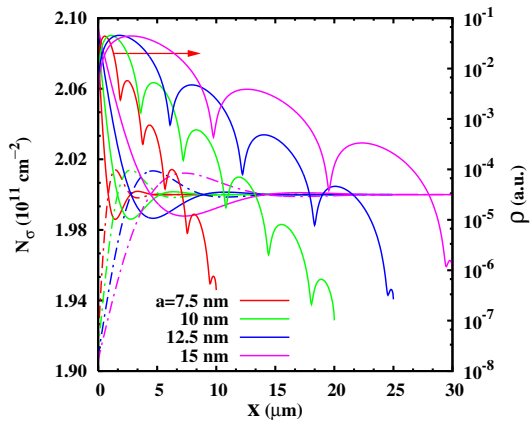,width=7.cm}}
\caption{(Color online)  $N_{\sigma}$ and  $\rho$
  {\it vs.} the position $x$ at different quantum well widths
  $a=7.5$, 10, 12.5, and 15\ nm (with $L=10$ ,20, 25 and 30\ $\mu$m
respectively) at $T=120$\ K.
The solid curves stand for $N_{\uparrow}$ and $\rho$,
 and  the dash-dotted ones represent $N_{\downarrow}$.
Note the scale of the incoherently summed spin coherence is
on the right hand side of the figure.
}
\label{fig:width}
\end{figure}

We further study the well-width dependence of the spin diffusion in the
steady state in Fig.\ \ref{fig:width},
with only the intrinsic scattering included at
$T=120$\ K. One finds from the figure that  both the
spin oscillation period and the spin diffusion length increase
markedly with the well width. This is consistent with the
spin R/D time in the time domain\cite{dephasing}
and is understood due to the decrease of the
 Dresselhaus spin-orbit coupling
Eq.\ (\ref{hk}) in which the linear term is proportional to  $1/a^2$.

\subsection{Magnetic field dependence of spin diffusion}

We investigate the effect of magnetic field on the spin R/D
in the steady state spin diffusion. It has been pointed out
by Weng and Wu that the magnetic
field plays a different role in spin diffusion/transport from the
time-domain spin precession in the spacial uniform case.\cite{PRB_weng1}
In the spin precession in the time domain, the magnetic field
alone without the DP term will not cause any spin R/D
due to the absence of inhomogeneous broadening.\cite{wu1} Whereas in
spin diffusion/transport, as the inhomogeneous broadening is caused
by $\Omega_{\bf k}$ which reads $m^\ast g\mu_B
B/k_x$ in the absence of the DP term
and still provides an inhomogeneous broadening leading to the
spin R/D.\cite{PRB_weng1} Furthermore, from
Eq.\ (\ref{omega}) one finds that the inhomogeneous broadening caused by the
applied magnetic field depends on the direction of the
magnetic field when it is combined with the inhomogeneous broadening
caused by the effective magnetic field $m^\ast{\bf h}({\bf k})/k_x$.
Therefore different direction of the magnetic field yields different spin
diffusion length.

\begin{figure}[htp]
\centerline{\psfig{file=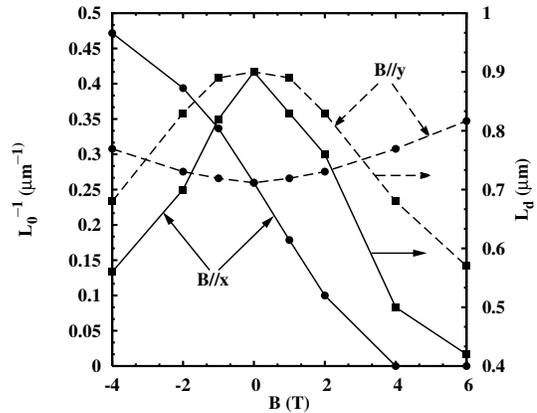,width=7.cm}}
  \caption{Inverse of the period of the spin oscillation $L_0^{-1}$
(curves with $\bullet$)
    and the spin diffusion length $L_d$
(curves with $\blacksquare$)  {\it vs}. the external magnetic
   field $B$  which is applied either vertical ($\| y$, dashed curves)
or parallel ($\| x$, solid curves) to the diffusion
    direction. $T=120$\ K and $N_i=0$.
    The dashed curves are for the vertical magnetic field and the
    solid curves are for the parallel one.
Note the scale of the diffusion length $L_d$ is
on the right hand side of the figure. }
  \label{fig:B}
\end{figure}

We reveal the magnetic field effect on the steady-state spin diffusion
by plotting in Fig.\ \ref{fig:B}
the spin diffusion length $L_d$ and the spin oscillation
period $L_0$, fitted from the calculated spin polarization along the
$z$-direction by  $\Delta
N_z(x)/N_e=C\exp(-x/L_d)\cos(2\pi x/L_0+\phi)$,
as function of the applied magnetic field  at $T=120$\ K.
The direction of the magnetic field is either
vertical (along the $y$-axis)  or parallel
(along the $x$-axis) to the diffusion direction.

One finds from the figure that the magnetic field
leads to additional spin R/D and therefore the spin diffusion length
decreases with the magnetic field, regardless of the direction of the
magnetic field. However, it is interesting to see that the
spin diffusion length has different symmetry when the magnetic field is
vertical or parallel to the spin diffusion direction: When
${\bf B}$ is parallel to the $y$-axis, $L_d(B)=L_d(-B)$; However,
when ${\bf B}$ is parallel to the $x$-axis, $L_d(B)\not=L_d(-B)$.
This can be understood from the fact  that the density matrices
from the kinetic spin Bloch equations\ (\ref{eq:BEQ})
have the symmetry $\rho_{k_x,k_y,kz}(B)=\rho_{k_x,-k_y,k_z}(-B)$
when ${\bf B}$ is along the $y$-axis. This symmetry is
broken if ${\bf B}$ is along the $x$-axis.

Differing from the magnetic field dependence of the spin diffusion length,
the spin oscillation period $L_0$  decreases monotonically with the
vertical magnetic field, regardless the
direction of the field. However $L_0$ increases with
the parallel magnetic field when the field is along the diffusion direction
and less than 4\ T (the spin
precession disappears when the field is larger than 4\ T)
but $L_0$ increases with the magnetic field when it is
anti-parallel to the diffusion direction. The spacial period is
determined by  $\mathbf \Omega_{0}+
m^\ast g\mu_B{\bf B}/\langle k_x\rangle$ with $\mathbf \Omega_{0}$ being
along the $x$-axis. $\langle k_x\rangle$ represents
 the average of $k_x$ due to the spin gradient and can be roughly estimated by
$\langle k_x\rangle=\sum_{\bf k}k_x\Delta f_{\bf k}/\sum_{\bf k}\Delta f_{\bf k}$,
which is a positive value due to the presence of
the spin gradient. For the vertical
magnetic field,  $\mathbf \Omega_{0}$ and  ${\bf B}/\langle k_x\rangle$
are perpendicular to each other, so the magnitude
of the spin oscillation period
is determined by $[\sqrt{(m^\ast g\mu_B{\mathbf B})^2/\langle k_x\rangle^2
+{\mathbf \Omega}_{0}^2}]^{-1}$ which always decreases with the
magnetic field. However, for the
parallel one, these two vectors are in the same direction and
the period is determined by $|\gamma(\langle k_y^2\rangle-\langle
k_z^2\rangle)+g\mu_{\mathtt{B}}B/\langle k_x\rangle|^{-1}$.
So the period increases with the
magnetic field. Moreover, it is interesting to see that when $B\sim$\ 4\ T,
 $|\gamma(\langle k_y^2\rangle-\langle
k_z^2\rangle)+g\mu_{\mathtt{B}}B/\langle k_x\rangle|\sim0$ and
further increasing of the magnetic field
does not lead to any spin oscillation due to the
too large  inhomogeneous broadening caused by the magnetic field.

\subsection{Electric field dependence of spin diffusion}

We now turn to study the electric field effect on spin
transport. An external electric field is applied along the $x$-axis.
When the electric field is large enough, the spin transport is then
in the hot-electron regime. It has been shown in our previous
investigation\cite{hot,strain} in the spacial homogeneous case
that the spin precession in the time domain is markedly affected by the
electric field. Especially an effective magnetic field
$\gamma v_d\pi^2/(a^2m^\ast)$,
which is proportional to the electric field,
is induced by the electric field combined with the
DP term due to the center-of-mass drift velocity $v_d=\mu E$
(and hence $k_x=v_d/m^\ast$) driven by the
electric field,
with $\mu$ representing the mobility.
This effective magnetic field leads to a spin precession
in the absence of any magnetic field and the precession
frequency changes with the
variation of the electric field.\cite{hot}
One may therefore expect that now in spin
transport, the spacial spin oscillation period is also proportional to the
electric field. This is {\em not} the case  as again $k_x$ in the
spin diffusion terms, {\em i.e.}, the third terms
in Eq.\ (\ref{eq:BEQ}), plays an
important role in the spin diffusion/transport and consequently the spacial
oscillation period $\Omega_0^{-1}$ is {\em independent} on $k_x$.
Therefore,  in spin transport the electric field cannot induce an
additional spacial spin precession.
It is further expected that in the steady state, $\langle k_x\rangle$ is
now determined by both the drift velocity and the spin gradient. Therefore,
if the electric field $E_x>0$ ($<0$), $\langle k_x\rangle$ is reduced
(enhanced) by the electric field and the spin diffusion length depends
on the direction of the electric field.

\begin{figure}[htp]
\centerline{\psfig{file=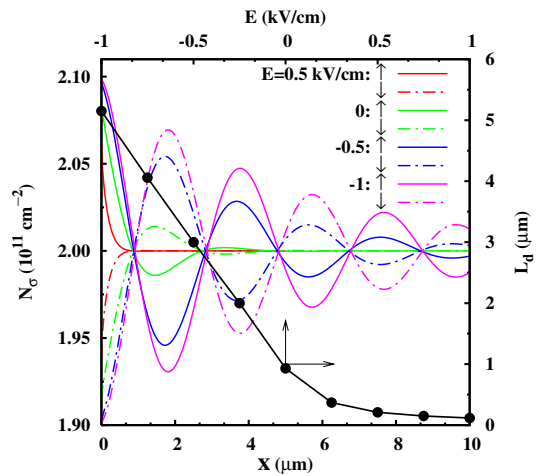,width=7.cm}}
\caption{(Color online) Spin-resolved electron density $N_{\sigma}$  {\it vs.} position
  $x$ at different electric field
  $E=0.5$, 0, $-0.5$ and $-1$\ kV/cm and the spin diffusion length $L_d$ against the
electric field $E$ at $T=120$\ K. $N_i=0$. It is noted that although
$x$ is plotted up to 10\ $\mu$m, $L=25$\ $\mu$m when $E=-1.0$ and $-0.75$\ kV/cm;
20\ $\mu$m when $E=-0.5$ and $-0.25$\ kV/cm; 10\ $\mu$m when $E=0$;
and 5\ $\mu$m when $E=0.5$, 0.75 and 1.0\ kV/cm.
 Note the scales of the spin diffusion length are on the top and
the right hand side of the figure.}
\label{fig:E}
\end{figure}

In Fig.\ \ref{fig:E} the spin density $\Delta N$ in the steady state
is plotted against the
position $x$ at different electric fields $E=0.5$, 0, $-0.5$ and $-1$\ kV/cm with
only the intrinsic scattering included at $T=120$\ K.
It is seen from the figure that the spin oscillation period
is almost unchanged when the electric field varies from
$-1$\ kV/cm to 0.5\ kV/cm.
This is consistent with the latest experimental observation in bulk
system.\cite{beck}
It is further noticed from the figure that the spin diffusion length
is markedly affected by the electric field. To reveal this effect, the
spin diffusion length is plotted as function of electric field in the same
figure. One finds when the electric field varies from $-1$\ kV/cm to $+1$\ kV/cm, the
spin diffusion length decreases monotonically. This can be easily understood
from the fact that when an electric field is applied along $-x$-direction ($+x$-direction), the
drift velocity caused by the electric field is along the $x$-direction
($-x$-direction), which
enhances (cancels) the velocity driven by the spin gradient. Therefore,
$\langle k_x\rangle$
is increased (reduced) and the inhomogeneous broadening is decreased
(enhanced). A longer (shorter) spin diffusion length is
then observed.

\section{Summary}
In summary, we have performed a theoretical investigation on the spin
diffusion/transport in $n$-type GaAs QWs at high temperatures ($\ge120$\ K)
by setting up and numerically solving the kinetic spin Bloch equations
together with the Poisson equation, with all the
scattering explicitly included. A new numerical scheme is developed
which can solve the kinetic equations with high accuracy and
speed.

The spin R/D mechanism in the spin diffusion/transport
is due to the inhomogeneous broadening, combined with the spin
 conserving scattering.
Differing from the spin precession in the spacial uniform system where the
inhomogeneous broadening is caused only by the momentum-dependent effective
magnetic field due to the SOC, in spin diffusion/transport, the inhomogeneous
broadening is caused additionally by the momentum
dependence in the spin diffusion
rate and therefore even the magnetic field alone can lead to spin R/D. Due to
the joint effects from the momentum-dependence in the diffusion rate and the
momentum dependence in the Dresselhaus effective magnetic field, there are
spin oscillations along the spin diffusion even in the absence of the external
magnetic field. The period of these spacial spin oscillations can be
affected by the temperature, the well width and the electron density,
 but is independent on the external electric field.
It increases when an additional scattering is added or the temperature/the well width
is increased.  Moreover, it is shown
that when the scattering is strong enough, the spin oscillations of electrons
at different momentum ${\bf k}$ show the {\em same} period. In fact, the intrinsic
scattering, {\em i.e.}, the electron-electron and the electron--LO-phonon scattering
is already strong enough to lead to this effect. This indicates that during the
spin diffusion, there is no interference-induced spin dephasing and $L_p=L_p^{\ast}$.

It is shown that the Coulomb scattering makes marked contribution to the spin
R/D in the spin diffusion/transport. It is
especially important in the hot-electron case as it provides the important
 mechanism of thermalization.
The scattering, the temperature, the QW width, the
magnetic field and the electric field dependence of the spin
diffusion length and the spacial spin oscillation period are
explored in detail. It is shown that in the weak scattering
regime, the spin diffusion length decreases if a new
scattering is added into the system, while in the strong scattering
regime, it increases.  In the
temperature regime we are interested in, the spin diffusion length
decreases with the temperature. Moreover, it increases with the increase of
the QW width as the inhomogeneous broadening decreases with the well
width but decreases with the external magnetic field
as the later adds a new inhomogeneous broadening into the system. The
external electric field can effectively prolong or suppress the spin
diffusion length depending on whether the electric field is
antiparallel or parallel to the spin diffusion direction.
We believe this investigation is useful for the understanding of the
spin transport as well as the design of the spintronic device.

\begin{acknowledgments}

This work was supported by the Natural Science Foundation of China
under Grant Nos.\ 90303012 and 10574120, the Natural Science Foundation
of Anhui Province under Grant No.\ 050460203, the
National Basic Research Program of China under Grant
No.\ 2006CB922005, the Knowledge Innovation
Project of Chinese Academy of Sciences and SRFDP. The authors acknowledge
valuable discussions with M. Q. Weng and M. P. Zhang.

\end{acknowledgments}

\appendix

\section{Numerical scheme for kinetic spin Bloch equations}
\label{method}
The kinetic spin Bloch
equations Eq.\ (\ref{eq:BEQ}) can be rewritten as:
\begin{eqnarray}
\frac{\partial}{\partial t}\rho_{\mathbf k}(x, t) +
\frac{k_x}{m^{\ast}}\frac{\partial}{\partial x}\rho_{\mathbf
  k}(x,t) = G_{\mathbf k}[x, t]
\label{wave}
\end{eqnarray}
with
$G_{\mathbf k}[x, t] =\left.\frac{\partial\rho_{\mathbf k}(x, t)}{\partial
    t}\right|_{\mathtt{dr}}
+ \left.\frac{\partial\rho_{\mathbf k}(x, t)}{\partial
    t}\right|_{\mathtt{coh}}
+ \left.\frac{\partial\rho_{\mathbf k}(x,t)}{\partial
    t}\right|_{\mathtt{scat}}$.
The discretization of the time and spatial derivatives
in Eq.\  (\ref{wave}) reads
\begin{widetext}
\begin{eqnarray}
&&\frac{1}{2}\left[\frac{\rho_{\mathbf k}(x, t+\Delta t)-\rho_{\mathbf
      k}(x, t)}{\Delta t} + \frac{\rho_{\mathbf k}(x+\Delta x, t+\Delta t)-\rho_{\mathbf
      k}(x+\Delta x, t)}{\Delta t}\right]+
\frac{k_x}{m^{\ast}}\frac{\rho_{\mathbf k}(x+\Delta x, t+\Delta t) -
  \rho_{\mathbf k}(x,t+\Delta t)}{\Delta x} \nonumber\\
&& =
\frac{1}{2}\left(G_{\mathbf k}[x ,
  t]+G_{\mathbf k}[x+\Delta x, t] \right)\ .
\label{dis}
\end{eqnarray}
For $k_x>0$, the electron is propagated from its left positions to the
right ones. Therefore, from Eq.\ (\ref{dis}), the iterative format is
\begin{eqnarray}
\rho_{\mathbf k}(x+\Delta x, t+\Delta t)=-\frac{1-r}{1+r}\rho_{\mathbf
  k}(x, t+\Delta t)+\frac{1}{1+r}(\rho_{\mathbf k}(x, t)+\rho_{\mathbf
  k}(x + \Delta x, t))+\frac{\Delta t}{1+r}(G_{\mathbf k}[x ,
  t]+G_{\mathbf k}[x+\Delta x, t])\ .
\end{eqnarray}
For $k_x<0$, the electron state is propagated from its right positions and
the iterative format is then
\begin{eqnarray}
\rho_{\mathbf k}(x, t+\Delta t)=-\frac{1-r}{1+r}\rho_{\mathbf
  k}(x+\Delta x, t+\Delta t)+\frac{1}{1+r}(\rho_{\mathbf k}(x + \Delta
x,t)+\rho_{\mathbf
  k}(x, t)+\frac{\Delta t}{1+r}(G_{\mathbf k}[x ,
  t]+G_{\mathbf k}[x+\Delta x, t])\ .
\end{eqnarray}
\end{widetext}
In these equations $r=\frac{2|k_x|\Delta t}{m^{\ast}\Delta x}$.

We truncate the spacial variable $x$ from 0 to $L$. At the left edge, the
boundary conditions are given only for the states with $k_x>0$
and at the right edge
only for those with $k_x<0$. Note the state with $k_x=0$ does not
contribute to the diffusion process but contributes to the local
source term $G_{\mathbf k}[x, t]$. With the combined effects of the
spin-flip terms and the scattering terms, the spin
signal decays when it diffuses into the QW in the scale
of the diffusion length $L_d$. When $L\gg L_d$, one can
study the spin injection properties.

The computation is carried out in a parallel manner in
the ``Beowulf'' cluster.
For a typical calculation with the partitions $16\times18$ grid points in the
$(k,\theta)$-space and $400$ points in the real space, it takes about 7 hours
for the system to evolute  to 300\ ps with a time step of 0.023\ ps by  6-node
 AMD Athlon XP3000+CPU's.


\begin{thebibliography}{0}
\bibitem{converg} We also checked the division
with $(N, M)=(32,34)$ and
$800$ grids in the real space and obtained the almost identical
results. So the accuracy of the numerical calculation is high
enough. The reason that for such a
rough ${\bf k}$-mesh the results are converged
lies on two facts: one is that, as shown in
Sec.\ III(A), all the density matrix elements in the ${\bf k}$-space
are in the same phase; the other is that the system is
almost in the non-degenerate regime. Therefore the spin
polarization in the $k$-space is extended to a very large energy scale
and is therefore very smooth.
\end{thebibliography}
\end{document}